\documentclass{osa-article}

\journal{oe}

\usepackage{graphicx}
\usepackage{dcolumn}
\usepackage{bm}
\usepackage{enumerate}
\usepackage{subfigure}
\usepackage{amsmath}
\usepackage{color}
\usepackage{lineno} 

\begin{document}

\title{Optical diffraction neural networks assisted computational ghost imaging through dynamic scattering media}

\author{Yue-Gang Li\authormark{1}, Ze Zheng \authormark{1}, Jun-jie Wang \authormark{1}, Ming He \authormark{4}, Jianping Fan \authormark{4}, Tailong Xiao \authormark{1,2,3,*}, and Guihua Zeng\authormark{1,2,3,\dag}}

\address{\authormark{1} State Key Laboratory of Photonics and Communications, Institute for Quantum Sensing and Information Processing, Shanghai Jiao Tong University, Shanghai 200240, P.R. China\\
\authormark{2}Hefei National Laboratory, Hefei, 230088, P.R. China\\
\authormark{3}Shanghai Research Center for Quantum Sciences, Shanghai, 201315, P.R. China\\
\authormark{4} AI Lab, Lenovo Research, Beijing 100094, P.R. China \\
}

\email{\authormark{*}tailong\_shaw@sjtu.edu.cn} 
\email{\authormark{\dag}ghzeng@sjtu.edu.cn}


\begin{abstract}
Ghost imaging leverages a single-pixel detector with no spatial resolution to acquire object echo intensity signals, which are correlated with illumination patterns to reconstruct an image. This architecture inherently mitigates scattering interference between the object and the detector but sensitive to scattering between the light source and the object. To address this challenge, we propose an  optical diffraction neural networks (ODNNs) assisted ghost imaging method for imaging through dynamic scattering media. In our scheme, a set of fixed ODNNs, trained on simulated datasets, is incorporated into the experimental optical path to actively correct random distortions induced by dynamic scattering media. Experimental validation using rotating single-layer and double-layer ground glass confirms the feasibility and effectiveness of our approach. 
Furthermore, our scheme can also be combined with physics-prior-based reconstruction algorithms, enabling high-quality imaging under undersampled conditions. 
This work demonstrates a novel strategy for imaging through dynamic scattering media, which can be extended to other imaging systems. 
\end{abstract}

\section{Introduction}

{\color{blue}Ghost imaging (GI) is an imaging technique that reconstructs an object using a single-pixel or bucket detector, and it can operate in a lensless configuration \cite{edgar2019principles}.} 
This imaging framework makes it particularly promising for applications in spectral ranges where array detectors or suitable lenses are difficult to implement, such as X-rays \cite{cheng2004incoherent, yu2016fourier, pelliccia2016experimental}, mid-to-far infrared \cite{radwell2014single}, and terahertz waves \cite{shrekenhamer2013terahertz}. Initially, GI is demonstrated through two-photon coincidence measurement based on quantum entanglement \cite{pittman1995optical}. Later, researchers discovered that it could also be achieved using classical pseudo-thermal or thermal light sources \cite{bennink2002two, valencia2005two}. As research progressed, the concept of computational ghost imaging \cite{shapiro2008computational, bromberg2009ghost} is introduced, replacing the reference beam with predefined illumination patterns, enabling imaging with only a single-pixel detector.  
Both single-arm computational ghost imaging and dual-arm ghost imaging share a fundamental requirement: knowledge of the illumination intensity pattern on the object plane. 

However, turbulence and scattering in the optical path can distort the illumination pattern on the object plane, breaking its correlation with the reference pattern, thereby leading to failure of the object reconstruction \cite{yang2015lensless}. To address this challenge, researchers have explored various approaches to enhance image reconstruction quality \cite{gao2020imaging, nie2021noise, lin2023scattering,liu2021self, paniagua2019blind, lin2022ghost, tananyan2024reciprocity, yuan2022unsighted, jauregui2019single, li2020compressive, gao2020computational, gao2022extendible, wang2019learning, lu2024multi, zhang2023imaging}, including improved illumination patterns \cite{gao2020imaging, nie2021noise, lin2023scattering,liu2021self}, spatial correlations between reflected \cite{paniagua2019blind} or transmitted \cite{lin2022ghost} patterns and object illumination patterns \cite{tananyan2024reciprocity, yuan2022unsighted}, filtering techniques \cite{jauregui2019single}, polarization information \cite{lu2024multi}, and deep learning \cite{li2020compressive, gao2020computational, gao2022extendible, wang2019learning}. 
Among these techniques, some are effective only for static scattering media, others are constrained by the optical memory effect (OME), and some depend primarily on ballistic photons. Therefore, developing an imaging approach that resists dynamic scattering remains a challenging yet crucial task.

Leveraging the intrinsic parallelism, multitasking capability, and low-latency nature of light, optical diffraction neural networks \cite{sciencea2018} have emerged as a highly stable and robust computational framework with remarkable efficiency and accuracy. In contrast to conventional electronic computing, ODNNs process information through the diffraction of light to accomplish specific tasks, and can be physically implemented using 3D printing \cite{luo2022computational, LAM2023010005}, metasurfaces \cite{chen2021diffractive}, or spatial light modulators \cite{sciadvadn2024zhang}. 
Recently, optical diffraction neural networks have been employed to mitigate the effects of optical scattering. The earliest demonstrations is conducted by the Ozcan group, who designed a THz-band ODNN-based imaging system capable of reconstructing both amplitude and phase objects behind scattering media. However, the scattering medium in their experiments was generated through simulation and subsequently fabricated via 3D printing, and thus the applicability of this approach to real, naturally occurring scattering conditions remains uncertain. 
Later, Gu’s group designed optical convolutional neural networks \cite{sciadvadn2024zhang} and compact ODNNs \cite{NP2025gumin} to counter scattering effects from scattering medium and multimode fiber, but these systems are limited to static scenes and do not generalize well to dynamic cases. In practical applications, scattering media like haze, turbulence, and biological tissue change over time. Thus, developing imaging methods that can handle dynamic scattering is crucial and highly applicable \cite{li2025realtimeimagingdynamicscattering}.

In this work, we propose an optical diffraction neural networks assisted computational ghost imaging scheme, which is dedicated to addressing the interference of the dynamic scattering medium between the light source and the object. This method integrates ODNNs into the optical path to correct dynamic scattering disturbances in real time, preserving the correlation between the illumination pattern on the object and the reference pattern. 
The ODNNs are trained entirely on simulated datasets without incorporating any prior knowledge of the experimental scattering medium. We experimentally demonstrate the effectiveness of our method with rotating single-layer and double-layer ground glass as dynamic scattering media. By integrating an untrained reconstruction algorithm into our framework, we achieve high-quality image recovery even under under-sampled conditions. This plug-and-play strategy is also applicable to other imaging frameworks.

\section{Materials and Methods}

\subsection{Model establishment and ODNN training }

\begin{figure}[h]
	\centering
	\includegraphics[scale=1.9]{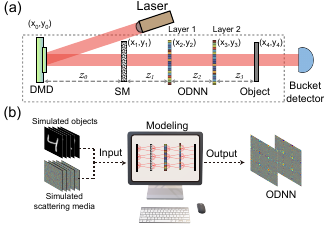}
	\caption{\label{fig:compare} (a) Schematic of scattering-robust computational ghost imaging assisted by a two layers optical diffraction neural network. 
    (b) Schematic diagram of ODNN training. First, the light field propagation from the DMD plane to the object plane is modeled. Subsequently, N different simulated objects and M different simulated scattering media are used as inputs to train a set of ODNNs. SM: Scattering Medium; DMD: Digital Micromirror Device. 
    }
\end{figure}

The experimental setup of our scheme is illustrated in Fig. \ref{fig:compare}(a). It builds upon a conventional computational ghost imaging system, assisted with an ODNN module. The ODNN is designed to correct the illumination patterns at the object plane, maintaining their correlation with the DMD-loaded patterns and thus mitigating the effects of scattering. To develop an ODNN capable of this functionality, we systematically address three key aspects in simulation: physical modeling, dataset generation, and training strategy.

We first extract the optical path segment from the {\color{blue}digital micromirror device} (DMD) plane to the object plane in Fig. \ref{fig:compare}(a) (dashed box) and construct a forward light field propagation model in computer (see Supplementary Note 2 for details). 
The objective of constructing is to optimize a set of ODNNs, i.e., a set of phase patterns distributed {$\phi_\mathrm{1}(x_2,y_2)$ and $\phi_\mathrm{2}(x_3,y_3)$} on the interval (0, 2$\pi$] for modulating the phase of the light field, by using simulated objects and simulated scattering media as inputs. As shown in Fig. \ref{fig:compare}(b).

In the ODNN architecture, adjacent layers are interconnected through optical diffraction. The maximum half-cone diffraction angle is described by ${\varphi _{\max }} = \arcsin \left( {\frac{\lambda }{{2\delta }}} \right)$ \cite{sciencea2018}, where $\lambda$ denotes the wavelength and $\delta$ represents the feature size of the neurons. 
In our work, $\lambda$ = 780 nm and $\delta$ = 25 $\mu$m, yielding $\varphi _{\max } =0.9^ \circ$. 
To ensure a sufficient number of neural connections, the distances between the layer 1 and layer 2 are 20 cm. 
Under the paraxial approximation, the Fresnel diffraction equation \cite{goodman2005introduction} well describes the free-space propagation between adjacent planes
\begin{equation}
\begin{aligned}
	U\left( {u,v} \right) = \frac{{{e^{ikz}}}}{{i\lambda z}} \iint U\left( {x,y} \right) \exp \left\{ {i\frac{k}{{2z}}\left[ {{{\left( {x - u} \right)}^2} + {{\left( {y - v} \right)}^2}} \right]} \right\}  \mathrm{d}x\mathrm{d}y,
	\label{eq1}
\end{aligned}
\end{equation}
where $k=\frac{2\pi}{\lambda}$ is the wave vector and $z$ is the propagation distance. $U\left( {u,v} \right)$ and $U\left( {x,y} \right)$ represent the light fields on two successive planes. For simplicity, we denote Fresnel propagation by ${\Im _z}\left\{  \cdot  \right\}$, allowing Eq. (\ref{eq1}) to be rewritten as $U\left( {u,v} \right) = {\Im _z}\left\{ {U\left( {x,y} \right)} \right\}$.
Considering a physical propagation model incorporating a two-layer ODNN, the system’s amplitude point spread function after one complete propagation is given by: 
\begin{equation}
\begin{aligned}
\mathcal{F}(x_4,y_4)&=\Im_{z_3}\{\Im_{z_2}\{\Im_{z_1}\{e^{ik\sqrt{z_0^2 + x_1^2 + y_1^2}}e^{i\phi(x_1,y_1)}\} e^{i\phi_{\mathrm{1}}(x_2,y_2)}\} e^{i\phi_{\mathrm{2}}(x_3,y_3)}\}
\end{aligned}
\label{eq2}
\end{equation}
where subscripts 0, 1, 2, 3, and 4 denote the coordinates of the DMD, scattering medium, layer 1, layer 2, and object plane respectively.  $z_0$, $z_1$, $z_2$, $z_3$, represent the propagation distances. 
$\phi(x_1, y_1)$ denotes the random phase introduced by the simulated scattering medium. 
$\phi_\mathrm{1}(x_2,y_2)$ and $\phi_\mathrm{2}(x_3,y_3)$ correspond to the phase patterns introduced by diffraction layer 1 and  layer 2, respectively. 
Without ODNN correction, i.e., for $\phi_\mathrm{1}(x_2,y_2)$ and $ \phi_\mathrm{2}(x_3,y_3)$ are constants, the point spread function, $|\mathcal{F}(x_4,y_4)|^2$, is a random speckle pattern. 
The objective of training the ODNN is to optimize $\phi_\mathrm{1}(x_2,y_2)$ and $ \phi_\mathrm{2}(x_3,y_3)$ to produce a clear image at the object plane, which is achieved by minimizing the loss function $\mathcal{L}$ through supervised learning 
\begin{equation}
    \mathcal{L}\left( {{\phi_\mathrm{1}},{\phi_\mathrm{2}}} \right) = -\frac{1}{{M N }}\sum\limits_{i = 1}^{{{M}} } \sum_{j=1}^{{N}} {P\left( {{O_j},{{\left| {\sqrt {{O_j}}  \circledast \mathcal{F}}_i \right|}^2}} \right)},
	\label{eq3}
\end{equation}
where $O_j$ represents the ground truth of the $j$-th input pattern, derived from the MNIST \cite{lecun2002gradient} dataset.
$N$ and $M$ are the numbers of simulated objects and simulated scattering media in the training set respectively, $\mathcal{F}_i$ is the amplitude point spread function of the system under the $i$-th simulated scattering medium, $\circledast$ denotes convolution operation and $P(\cdot)$ is the Pearson correlation coefficient (PCC). Training details in Supplementary Note 3. 

The well-trained ODNNs can reconstruct input object images even in the presence of unknown objects and scattering media. In our previous work \cite{li2025realtimeimagingdynamicscattering}, we have demonstrated that this capability stems from the optical shower-curtain effect. Consequently, the ODNN trained solely on fully simulated data can also be applied to imaging through real scattering media, with effectiveness within 1–2 transport mean free paths.

\subsection{Image reconstruction algorithm.}

\begin{figure}[hb]
	\centering
	\includegraphics[width=0.9\textwidth]{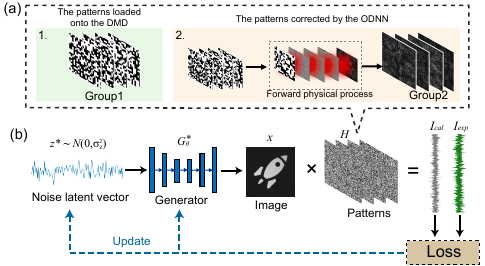}
	\caption{\label{fig:untrain} (a) Two groups of illumination patterns used for object image reconstruction. Group 1: patterns projected onto the DMD; Group 2: patterns of Group 1 after the forward physical process. (b) Schematic of image reconstruction performed by the untrained reconstruction algorithm. }
\end{figure}

During image reconstruction, the most straightforward strategy is to use the patterns loaded onto the DMD (Fig. \ref{fig:untrain}(a), Group 1). Although the ODNN is designed to ensure that these patterns closely match the actual illumination at the object plane, discrepancies still remain in practice. To minimize this mismatch, we also consider reconstructing the image using illumination patterns (Group 2) simulated via a forward optical propagation model. 

In our work, we consider two reconstruction algorithms. The first is differential ghost imaging (DGI) \cite{DGI2010}, an improved version of the conventional correlation-based method, defined as:

\begin{equation}
DGI(x,y) = \left\langle {{B_1}{I_2}\left( {x,y} \right)} \right\rangle  - \frac{{\left\langle {{B_1}} \right\rangle }}{{\left\langle {{B_2}} \right\rangle }}\left\langle {{B_1}{I_2}\left( {x,y} \right)} \right\rangle 
\label{eq4}  
\end{equation}
$B_1$ is the collected bucket detection signal, while $B_2 = \sum{{I_2}\left( {x,y} \right)} $ serves as the reference bucket detection signal. $I_2(x,y)$ represents the illumination pattern. Typically, the result obtained from Eq. \ref{eq4} is used as a benchmark. 

To further improve the reconstruction quality, we incorporate a variable generative network (VGenNet) \cite{zhang2023vgennet, brock2018large, xiao2024quantum} into our pipeline, as illustrated in Fig. \ref{fig:untrain}(b). 
VGenNet is a hybrid architecture that combines a generative network $G_\theta(z)$ with a physical model $H$. As it does not rely on large-scale training data to learn its parameters, it is also referred to as an untrained reconstruction algorithm. 
The reconstruction process of the untrained algorithm proceeds as follows. First, a generator $G_\theta(z)$ takes a random noise vector $z$ as input and generates an image  $x$. Next, the imaging model $H$ maps the generated image $x$ to a one-dimensional bucket signal $I_{cal}$, where the i-th element can be expressed as $I_{cal}^i = \sum\limits_{u,v} {H(u,v,i)x(u,v)}$. The error between $I_{cal}$ and the measured intensity sequence $I_{exp}$ is then computed and used to update both the latent vector $z$ and the generator parameters  $G_\theta(z)$ via backpropagation. As optimization proceeds, the generator output is expected to converge to the desired image. 
The reconstruction process of this algorithm can be formulated as: 

\begin{equation}
{z^*},G_\theta ^ * = \mathop {\arg \min }\limits_{z,{G_\theta }} \frac{1}{n} \left\| {I_{exp} - H{G_\theta }(z)} \right\|_2^2+TV({G_\theta }(z)),~~~s.t.~~{x^*} = G_\theta ^ *({z^ *})
\label{eq5}   
\end{equation}
The first term represents the mean squared error (MSE), where $n$ denotes the dimension of bucket detection, and the second term represents the total variation (TV) regularization. 
It is important to note that a critical requirement of the untrained algorithm is that the patterns used to compute $I_{cal}$ must closely approximate those that illuminate the object in the actual experiment. This ensures the accuracy of the physical model, allowing the loss to primarily reflect the quality of the generator’s output. This implicitly demands that the ODNN be precisely calibrated.

\section{Results}

\begin{figure}[hb]
	\centering
	\includegraphics[width=0.9\textwidth]{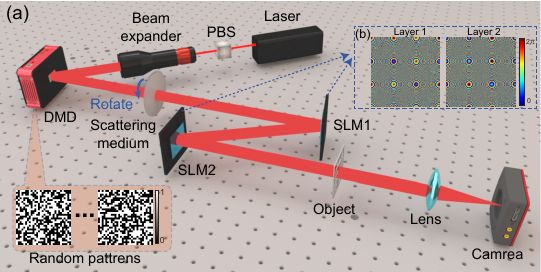}
	\caption{\label{fig:setup} (a) Schematic diagram of the experimental setup for the ODNNs-assisted ghost imaging. {\color{blue}PBS, Polarizing Beam Splitter; DMD, Digital Micromirror Device; SLM, Spatial Light Modulator}. (b) A two-layer ODNN trained with a simulated dataset.  }
\end{figure}

To validate the proposed imaging scheme, we built the experimental setup shown in Fig.\ref{fig:setup}(a). 
A 780 nm laser (DBR780PN, Thorlabs) passes through a polarizing beam splitter (PBS) and a 10X beam expander before being modulated by a digital micromirror device (DMD, VX4100, Vialux) to encode structured illumination patterns. The modulated beam traverses a rotating scattering medium (single or double layer 220-grid diffuser, Thorlabs) and an optical diffraction neural network before illuminating the object. A collecting lens and a camera (acA2448-75umBAS, Basler) capture the transmitted light intensity, serving as the bucket detection signal. In the experiment, ODNN is implemented with two phase-only spatial light modulators (SLM, X15213-02, Hamamatsu), each featuring a effective resolution of $256 \times 256$ pixels with a pixel size of 25 $\mu$m $\times$ 25 $\mu$m. 
The camera, triggered by the DMD, has an exposure time of 40 ms. The scattering medium is mounted on a motorized rotation stage, which yielded a speckle decorrelation time below 1 ms (Fig. \ref{fig:fullSampling}(g)). 

\subsection{Performance and evaluation.}

\begin{figure}[h]
	\centering
	\includegraphics[width=0.95\textwidth]{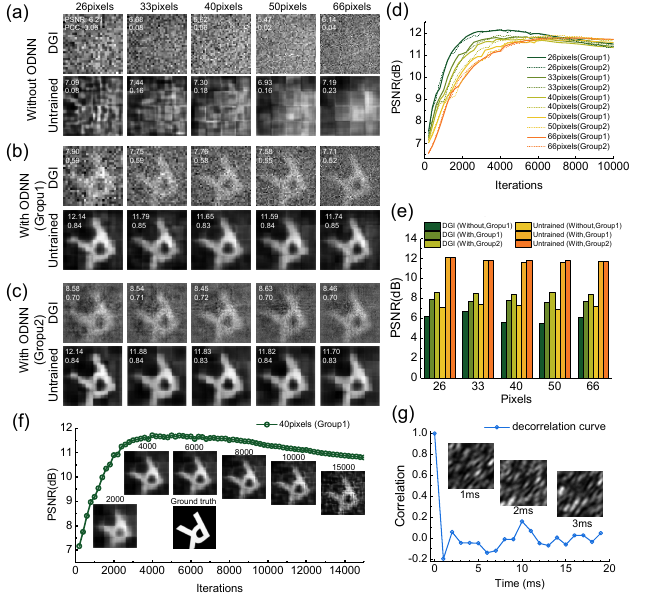}
	\caption{\label{fig:fullSampling} Image reconstruction results obtained using illumination patterns with different transverse coherence scales (sizes). (a) Reconstruction result without ODNN correction. (b) and (c) show the reconstructed results based on the patterns from Group 1 and Group 2, respectively, when ODNN correction is implemented. (d) The iteration process curves of the untrained algorithm results in (b) and (c). (e) Quantitative comparison of the results in (a), (b), and (c), with PSNR as an example. (f) Reconstruction results of the untrained algorithm under 40 $\times$ 40 pixels illumination pattern at different numbers of iterations. (g) Speckle decorrelation curve of a rotating 220-grid ground glass. }
\end{figure}

We first conducted experiments using illumination patterns with five different transverse coherence scales (sizes). The corresponding coherence lengths are 162 $\mu$m, 129.6 $\mu$m, 108 $\mu$m, 86.4 $\mu$m, and 64.8 $\mu$m, with illumination patterns sizes of 26 $\times$ 26, 33 $\times$ 33, 40 $\times$ 40, 50 $\times$ 50, and 66 $\times$ 66 pixels, respectively (as shown in Fig. S6). 
All reconstructions results in Fig. \ref{fig:fullSampling} are performed under full sampling conditions. That is, for an illumination pattern of 26 $\times$ 26 pixels, a total of 676 measurements are acquired to complete the reconstruction.  
In the case of without ODNN correction (Fig. \ref{fig:fullSampling}(a)),  object images could not be reconstructed regardless of using the DGI or untrained algorithms. This is attributed to the severe distortion of the illumination patterns on the object plane. In contrast, with ODNN correction, object reconstruction is achieved in all five cases (Fig. \ref{fig:fullSampling}(a) and \ref{fig:fullSampling}(b)).

In the DGI results, the reconstruction results using group 2 is slightly better than that using group 1 (Fig. \ref{fig:fullSampling}(e)), which is consistent with the expected results, as group 2 patterns share more similar features with the real illumination patterns. 
However, this advantage is not clearly retained in the reconstruction results of untrained algorithms.
This is because physics-driven model inherently incorporate more prior information, which mitigates the influence of illumination patterns on reconstruction quality, making the reconstruction results tend to be identical. 
This observation is further supported by the convergence curves shown in Fig. \ref{fig:fullSampling}(d). 
Notably, as the number of iterations increases, the curve in Fig. \ref{fig:fullSampling}(d) first rises and then gradually decreases. 
This degradation is caused by overfitting, which arises from the discrepancies between the actual illumination patterns and those used in the reconstruction. 
Since the 26 $\times$ 26 pixels and 33 $\times$ 33 pixels illumination pattern involve fewer bucket measurements, they exhibit overfitting earlier under the same number of iterations. 
When the number of pixels of illumination pattern increases, more iterations are usually required for overfitting to occur. Fig. \ref{fig:fullSampling}(f) presents the reconstruction results evolution over 15,000 iterations for the 40 $\times$ 40 pixels case.

\begin{figure}[h]
	\centering
	\includegraphics[width=0.9\textwidth]{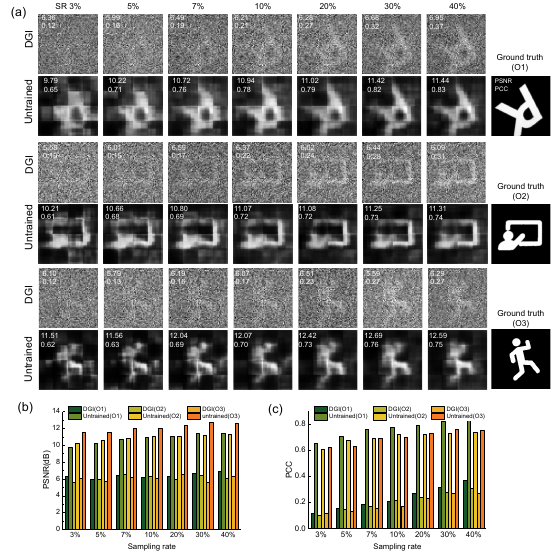}
	\caption{\label{fig:differentSampling} (a) Image reconstruction results of three objects at different sampling rates, where the illumination pattern size is 66×66 pixels. (b) and (c) show the PSNR and PCC metrics of the reconstruction results in (a), respectively. O1, O2, and O3 are abbreviations of object1, object2, and object3, respectively. }
\end{figure}

Furthermore, we take the 66 $\times$ 66 pixels illumination patterns as an example to analyze reconstruction performance under varying sampling ratios. The results are shown in Fig. \ref{fig:differentSampling}(a). 
When the sampling ratio is below 10\%, the limited useful information and the influence of noise lead to poor reconstruction results. In the DGI reconstruction, the object is barely discernible, and the untrained result also exhibits substantial deviation from the ground truth. 
Once the sampling ratio reaches 30\%, satisfactory reconstruction can be achieved. Figs. \ref{fig:differentSampling}(b) and \ref{fig:differentSampling}(c) show the peak signal-to-noise ratio (PSNR) and Pearson correlation coefficient of the reconstructed results, respectively.

When the exposure time of the camera (act as bucket detector) exceeds the speckle decorrelation time, the illumination pattern on the object plane is the intensity sum of patterns under the interference of multiple different scattering media. 
Considering this effect, we simulate this process by generating multiple simulated scattering media and summing the corrected intensity patterns obtained under each scattering medium to form the final Group 2 pattern. 
Fig. \ref{fig:differentSampling}(a) shows the reconstruction results using Group 2 patterns generated with 1, 2, 4, 8, 10, and 15 simulated scattering media, respectively. As the number of scattering media increases, the results recovered by DGI gradually improve. This trend indicates that multi-intensity patterns summation suppresses noise and enhances the accuracy of illumination patterns. 
For untrained algorithms, physical prior information significantly improves image quality. This makes minor improvements in the illumination pattern less noticeable for image quality, and the final reconstruction results under different conditions often tend to be identical.

\begin{figure}[h]
	\centering
	\includegraphics[width=0.95\textwidth]{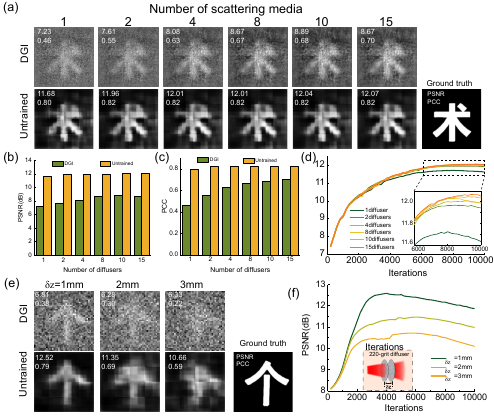}
	\caption{\label{fig:different_diffuser} (a) Image reconstruction results using Group 2 patterns obtained under different numbers of scattering media. The illumination pattern size is 40 $\times$ 40 pixels. (b) and (c) show the PSNR and PCC metrics of the reconstruction results in (a), respectively. (d) Iteration process curves of the untrained algorithm for different cases in (a). 
    (e) Reconstruction results under double-layer ground glass with different interlayer spacings, where the illumination pattern size is 40 $\times$ 40 pixels (Group 1). (f) Iteration process curves of the untrained algorithm for different cases in (e).}
\end{figure}

To evaluate the scattering tolerance limit of our system (or ODNN), we constructed a more challenging scattering medium by stacking two layers of 220-grit ground glass and increased the scattering strength by enlarging the interlayer spacing. The results are shown in Fig. \ref{fig:different_diffuser}(e). 
We observed a gradual decline in image quality with increasing interlayer distance, indicating a degradation in the correction capability of ODNN under stronger scattering conditions. 
One possible reason for this deterioration is the mismatch between the experimental configuration (i.e., a double-layer scattering medium) and the model assumption based on a single-layer phase screen. 
Notably, when we increased the number of simulated scattering layers to better reflect the experimental setup, the ODNN failed to converge during optimization. 

The reason for this degradation is that the optical field acquires a random phase distribution after passing through the first layer of ground glass, and subsequent propagation causes the optical fields to interfere with each other, leading to rapid decorrelation of the image. Typically, the thickness of the scattering medium is limited to 2–3 mm \cite{li2025realtimeimagingdynamicscattering}. These results indicate that implementing an ODNN robust to dynamic and strong scattering may be fundamentally challenging, potentially constrained by mathematical or physical limitations, which merits further research.

\section{Conclusion and Discussion}

In conclusion, we present an ODNN-assisted ghost imaging method for imaging through dynamic scattering media. 
The proposed scheme has three main advantages. 
First, the training of ODNN is completely based on simulation datasets and does not rely on real experimental data. Second, by jointly training on simulated random phase screens and multiple scattering conditions, we obtained a set of ODNN capable of generalizing to real scattering environments. 
Moreover, ODNN is not only unrestricted by dynamic scattering scenarios; on the contrary, it transforms dynamic scattering into a means to enhance the quality of reconstructed images. 
Finally, our imaging system can be combined with untrained algorithms to significantly enhance the quality of reconstructed images and operate at low sampling rates. 
We experimentally validated our method using rotating single- and double-layer ground glass diffusers. Our method offers ultrafast processing speed and holds promise for real-time distortion correction in a broad range of imaging systems.

However, our scheme also has limitations.
Experimental results show that our scheme exhibits strong robustness in the case of thin scattering, but the imaging results degrade significantly under multi-layer or volume scattering. 
This result indicates that obtaining a fixed modulation module robust to think dynamic scattering remains challenging.
Secondly, the imaging system needs to be built according to the structure of the simulation training, which may lack flexibility in real-world scenarios. 
In our model, the distance between the DMD and the scattering medium can be adjusted within a range of around ten to twenty centimeters. However, an increase in this spacing may lead to a reduction in the resolution of the illumination pattern on the object plane, which in turn causes degradation of the results. The scattering medium, diffraction layer, and object can be moved within a range of 3–4 cm.

Furthermore, we would like to emphasize that the fundamental reason for the feasibility of our scheme stems from the optical shower-curtain effect \cite{dror1998experimental, edrei2016optical}, a physical effect that utilizes the spatial correlation between the front and back sides of the scattering medium. 
This effect is usually effective within 1-2 transport mean free paths \cite{li2025realtimeimagingdynamicscattering}. 
The morphology of the diffraction layer are closely related to the specific tasks performed by the ODNN. In our work, the ODNN is trained to generate an upright, unit-magnification image; it can also generate scaled or rotated images, as shown in Fig S7, which demonstrates the flexibility of our method in imaging tasks. With the emergence of nonlinear optical diffraction neural networks \cite{wang2023image, zhang2025highly, zuo2019all, wu2025coupling}, we anticipate further improvements in anti-scattering capability and image quality.

\appendix


\section*{Acknowledgement}
This work was supported by the National Natural Science Foundation of China (No. 62401359), the fund of the State Key Laboratory of Photonics and Communications, the Innovation Program for Quantum Science and Technology (Grant No. 2021ZD0300703), Shanghai Municipal Science and Technology Major Project (2019SHZDZX01) and SJTULenovo Collaboration Project (202407SJTU01-LR019). 
\section*{Disclosures.} The authors declare no conflicts of interest.
\section*{Author Contributions.} G.Z. conceived the research project. T.X. designed the scheme. Y.L. constructed the theoretical model, operated the numerical simulation with assistance from J.W., carried out the experiment with assistance from Z.Z.. M.H. and J.F. provided assistance with the network training. 
All the authors discussed and contributed to the writing of the manuscript. 
\section*{Data availability.} Data underlying the results presented in this paper are not publicly available at this time but may be obtained from the authors upon reasonable request.

\bibliography{ref}

\end{document}